Calculations of the thermodynamic and kinetic properties of $LiV_3O_8$

Tonghu Jiang*, Michael L. Falk†

*†Department of Materials Science and Engineering, Johns Hopkins University, Baltimore, MD 21218 USA

†Department of Mechanical Engineering and Department of Physics and Astronomy, Johns Hopkins University, Baltimore, MD 21218 USA

Abstract

The phase behavior and kinetic pathways of $Li_{1+x}V_3O_8$ are investigated by means of density functional theory (DFT) and a cluster expansion (CE) methodology that approximates the system Hamiltonian in order to identify the lowest energy configurations. Although DFT calculations predict the correct ground state for a given composition, both GGA and LDA fail to obtain phase stability consistent with experiment due to strongly localized vanadium 3d electrons. A DFT+U methodology recovers the correct phase stability for an optimized U value of 3.0eV. GGA+U calculations with this value of U predict electronic structures that qualitatively agree with experiment. The resulting calculations indicate solid solution behavior from $LiV_3O_8$ to $Li_{2.5}V_3O_8$ and two-phase coexistence between $Li_{2.5}V_3O_8$ and $Li_4V_3O_8$. Analysis of the lithiation sequence from $LiV_3O_8$ to $Li_{2.5}V_3O_8$ reveals the mechanism by which lithium intercalation proceeds in this material. Calculations of lithium migration energies for different lithium concentrations and configurations provides insight into the relevant diffusion pathways and their relationship to structural properties.

I. INTRODUCTION

Lithium ion batteries continue to receive intense academic and industrial interest due to their greater energy density, both with respect to volume and mass, than traditional batteries.[1] Since the first lithium ion battery was invented in the 1990s, extensive research has been conducted to explore potential cathode, anode and electrolyte materials in order to improve battery performance.

Of numerous proposed lithium ion battery cathodes, some vanadium oxides have drawn interest, particularly $V_2O_5$, $V_6O_{13}$ and $Li_{1+x}V_3O_8$. Though nanostructured $V_2O_5$ has been shown to have fairly good performance,[2] $V_2O_5$ and $V_6O_{13}$ generally suffer from capacity loss and low lithium diffusivity.[3,4] On the other hand, the structure of $Li_{1+x}V_3O_8$, layered tri-vanadate, is similar to commercially successful $LiCoO_2$, which consists of transition metal oxide layers and has been reported to exhibit high capacity, high lithium diffusivity and a long life cycle.[5,6] Further, compared with other prospective cathode materials, $LiV_3O_8$ has the advantage of low cost, material abundance and relatively easy synthesis.[7,8]

It was not until 1981 that electrochemistry of $Li_{1+x}V_3O_8$ was studied by Murphy[9] in both crystalline and glassy form and subsequently high capacities and good cyclability was obtained by Pistoia.[10,11] Efforts were made to improve its electrochemical performance and strengthen its practical applicability. Replacing lithium fully or partially with Na or Mg has been found to improve performance.[12-15] The amorphous form of $Li_{1+x}V_3O_8$ was shown to have advantages over its crystalline form,[16] as more lithium can be intercalated and faster lithium diffusion can be achieved.[17] Oxygen deficient $Li_{1+x}V_3O_8$ was also found to hold more lithium.[18] Ultrasonic treatment of $Li_{1+x}V_3O_8$ improves both specific capacity and cyclability.[19] Recent experimental research has focused on facile and large-scale production.[7,8] In a demonstration of the potential for practical application, Li/ Li$_{1+x}$V$_3$O$_8$ cells have been assembled to build a 200V 2kWh multicell system.[20]

Since its introduction as a cathode material, a variety of characterization methods have been applied to study the crystal structure of this material: X-ray diffraction,[10,21] neutron diffraction,[22] FTIR and XAS,[17] IR and Raman.[23] However, there remains some disparity between the reported phase behavior with some investigators reporting[24,25] the onset of a two-phase process at Li$_{2.5}$V$_3$O$_8$, while others[21,26] reporting single phase behavior up to Li$_3$V$_3$O$_8$. Hence, it is still unclear from experiment whether the two-phase process starts at Li$_3$V$_3$O$_8$ or Li$_{2.5}$V$_3$O$_8$. (We denote the starting structure as Li$_{2.5\sim3}$V$_3$O$_8$ temporarily. ) X-ray diffraction studies to date have been unable to identify all the possible lithium sites within the structure.[21] Although recent neutron diffraction studies have identified additional lithium sites,[22] no information is currently available regarding structural transitions that occur during lithiation and delithiation. Thus the detailed nature of the two-phase process, the structural transformation that occurs from Li$_{2.5\sim3}$V$_3$O$_8$ to Li$_4$V$_3$O$_8$, remains an open question. Since experimental techniques have been limited in their ability to detect atomic scale details regarding the structure and kinetics, theoretical approaches can play a vital role in filling in the gaps in our understanding of these materials. The goal of this paper is to elucidate the above-mentioned issues by applying a number of computational techniques, which are introduced in section II. After reviewing the experimental and computational literature on these compounds in section III, we detail the results of our investigations in section IV using standard density functional theory (DFT) calculations and cluster expansion approaches. We further discuss discrepancies between these results and the experimental literature. In section V we apply the DFT+U methodology and show that this results in a more adequate prediction of the phase behavior. In section VI, we analyze the kinetic pathways for lithium diffusion through the LiV$_3$O$_8$ structure.

## II. METHODOLOGIES

Density functional theory (DFT) has proven useful for predicting the relative energies of solid state periodic systems and estimating defect formation and activation energies that control materials kinetic processes[27]. It is employed by material scientists to investigate electronic, mechanical, magnetic and other properties for a broad range of materials. Practical implementation of DFT requires that the exact exchange and correlation contribution to total energy be approximated. This is typically accomplished by applying the local density approximation (LDA) or the generalized gradient approximation (GGA). These are used in combination with a variety of pseudo-potentials that describe the effective potential experienced by the explicitly modeled valence electrons in the presence of the core electrons.

Despite successful application of DFT in numerous contexts, it fails to describe the electronic structure of strongly correlated materials such as transition metal oxides due to existence of localized d electrons. Attempts have been made to recover the correct insulating behavior of transition metal oxides NiO[28] and FeO[29] by developing the DFT+U methodology. In this approach, an on-site coulomb repulsion term in the d-electron bands is added to the DFT Hamiltonian. This repulsion favors that the d-orbital be fully occupied or empty. A number of calculations based on DFT+U have been demonstrated to correctly describe the electronic structure,[29] but other shortcomings can arise in the predicted material properties due to the fact that this approach is still essentially a single particle approximation.[30]   For example optimizing agreement with different experimentally determined values appears to require different choices of the U parameter.[31] The U value in this approach has to be determined either by a linear response approach[29] or by emprically fitting to physical properties[31].

In order to compute thermodynamic properties for a compound or alloy system one needs to be able to generate formation energies for any realizable configuration, or at least those which significantly contribute to the phase space sampled by the system at the temperature of interest. While it is, in principle, possible to calculate the energy for any given structural state using a DFT or DFT+U calculation, these calculations are too computationally costly to undertake by brute force. To overcome this limitation, well-established cluster expansion (CE) methods[35,32] were used to approximate the system Hamiltonian. The cluster expansion decomposes the system energy as a function of effective interactions and site-correlation functions expressed as a series expansion such as

$$E(\sigma) = J_0 + \sum_i J_i S_i(\sigma) + \sum_{j<i} J_{ij} S_i(\sigma) S_j(\sigma) + \sum_{k<j<i} J_{ijk} S_i(\sigma) S_j(\sigma) S_k(\sigma) + \cdots$$
(1)

where the state of the system, $\sigma$, depends on the configuration of lithium atoms and vacancies, which is denoted by a set of spin variables $S(\sigma)$. Each lattice site has a spin variable; it is +1 or -1 if occupied by a lithium atom or a vacancy respectively. The site correlation functions are products of spin variables of particular group of lattice sites called a cluster, which can denote a pair, triplet, and so forth. All clusters that are related by symmetry operations have the same effective interactions. Thus the number of unknown coefficients is greatly reduced for high symmetry systems. These coefficients are obtained by fitting formation energies from first principles calculations to the linearized cluster expansion given in Eq. (1) using a traditional genetic algorithm to obtain an optimized result.[33]

After obtaining an optimized cluster expansion from a subset of lithium and vacancy configurations, one can search for ground states by Monte Carlo calculation, brute force calculation or genetic algorithm. If a new ground state is predicted its energy is checked by DFT calculation and this new information is incorporated into the fitting process. This process is repeated until all ground states are properly predicted.

## III. THE $LI_{1+x}V_3O_8$ SYSTEM

The $Li_{1+x}V_3O_8$ crystal belongs to monoclinic system with space group $P2_1/m$ and is one member of hewettite group. The formula of $Li_{1+x}V_3O_8$ implies that the octahedrally coordinated lithium atoms cannot be extracted; this would cause the oxidation state of V to exceed +5. The structural unit (primitive cell) contains two $Li_{1+x}V_3O_8$ molecules. During lithium intercalation a plateau is evident in the voltage profile implying two-phase coexistence over a range of intermediate lithium concentrations. Adopting the notation of Benedek,[34] structurally distinct low and high-lithiated phases are referred to as $\gamma_a$ and $\gamma_b$ respectively.

The $\gamma_a$ phase is commonly described as a structure formed by sheets of vanadium-oxygen polyhedra, i.e, distorted octahedra ($VO_6$) and distorted trigonal bipyramids ($VO_5$). Lithium ions are believed to reside in interlayer sites on the b-c plane, and these ions are assumed to contribute to the cohesive energy that binds these sheets together. The distorted octahedra and trigonal bipyramids share edges to form zigzagged ribbons and single chains along the b axis, respectively. The ribbons and chains are connected by sharing corners to form puckered vanadium-oxygen host layers. Therefore, this material is often referred to as a layered structure.

The $\gamma_b$ phase is significantly more ordered than the $\gamma_a$ phase, forming a defected rock salt-like structure. Lithium ions are known to occupy all possible octahedral sites as lithium composition increases. As discussed in section I, the structural transition of Li$_{1+x}$V$_3$O$_8$ from $\gamma_a$ to $\gamma_b$ phase is not well understood. Nevertheless, recent data obtained from neutron-diffraction[22] provides some insight into the phase transition path from $\gamma_a$ to $\gamma_b$. Fig.1 shows all the available sites for lithium occupancy in the $\gamma_a$ phase where there is one octahedral site $Li_1^\alpha$ and three tetrahedral sites $Li_2^\alpha, Li_a^\alpha, Li_b^\alpha$. In the $\gamma_b$ phase there are five octahedral sites $Li_1^\beta$, $Li_2^\beta, Li_3^\beta, Li_4^\beta, Li_5^\beta$. A nearly one-to-one mapping exists between the two structures, the site to site mapping is between $Li_1^\alpha$ and ($Li_4^\beta, Li_5^\beta$), $Li_a^\alpha$ and $Li_3^\beta$, $Li_b^\alpha$ and $Li_1^\beta$, $Li_2^\alpha$ and $Li_2^\beta$. A transition from $\gamma_a$ to $\gamma_b$ involves lithium atoms in $Li_1^\alpha$ octahedral sites hopping to one of the neighboring octahedral sites $Li_4^\beta, Li_5^\beta$ while the lithium atoms in the other tetrahedral sites $Li_2^\alpha, Li_a^\alpha, Li_b^\alpha$ only need to adjust slightly to adopt octahedral symmetry becoming $Li_2^\beta, Li_3^\beta, Li_1^\beta$. Contradicting this picture, Benedek, *et al.* [34] found good agreement between experiments and DFT computation based on the lithium sites suggested in Ref.21, which we will refer to as set A. These differ from the neutron diffraction results reported above,[22] which we will refer to as set B. Both plausible sets of lithium sites were considered in our DFT calculations. We found set A to be unsuitable for reasons we will now discuss.

In the DFT study by Benedek *et al.*,[34] in order to reduce the underestimation of cell volume they impose the experimentally determined lattice constants of Li$_{1.2}$V$_3$O$_8$ for calculations regarding the $\gamma_a$ phase and further impose a set of optimized lattice constants for Li$_4$V$_3$O$_8$ when investigating the $\gamma_b$ phase. In calculations performed with a fixed lattice parameter it is possible that certain structures are artificially stabilized by the imposed boundary condition. If internal degrees of freedom are allowed to relax, structures that appear unstable given the constraint may converge to lower energy states with significantly different crystal structures. Such full relaxation of the crystal structure is commonly practiced and preferred since it admits the possibility of mechanically unstable structures.[35] In our calculations numerous cases of unstable lithium configurations were observed which would have been overlooked if full relaxation were not employed.

It is perhaps for this reason that this previous work[34] found the $\gamma_a$ phase to be stable up to $Li_2V_3O_8$. Full relaxation of lowest energy structures in Ref.34 indicates that $Li_1V_3O_8$, $Li_{1.5}V_3O_8$ are stable, but $Li_2V_3O_8$ leads to a sheared crystal structure with dramatic changes of lattice parameters. Tests of a number of other configurations based on set A show similarly strong instabilities. From these observations we believe that set A does not accurately represent the sites that lithium occupies in the $\gamma_a$ phase. The investigations that follow will focus on results obtained from calculations based on set B. Note that unstable states are also encountered when set B is used, however this phenomena is only related to instability of internal degrees of freedom. Lithium atoms may shift position, but cell parameters do not vary significantly.

## IV. DENSITY FUNCTIONAL THEORY CALCULATIONS

To analyze the phase stability of the $Li_{1+x}V_3O_8$ system DFT methods as implemented in the plane wave code VASP[36,37] were used to calculate the total energies of a wide range of structures. Both local density approximation (LDA)[38] and generalized gradient approximation (GGA)[39] were used to approximate the exchange correlation energy. Projector augmented wave[40,41] (PAW) pseudopotentials were used for Li, V, O. A kinetic energy cutoff of 520eV was used in all calculations for both LDA and

GGA. Energy convergence with respect to k-points was tested on several lithium configurations in both $\gamma_a$ and $\gamma_b$ phase. A 6×9×3 Monkhorst-Pack[42] k-point mesh was found to be sufficient to accurately calculate formation energies with error smaller than 10mV for a single primitive cell using either LDA or GGA.

Total energies were obtained by relaxing all atomic coordinates and cell parameters using a conjugate gradient method while maintaining the symmetries of the cell. The convergence condition was such that all forces on atoms were smaller than 0.01 eV/Å. Investigations of the effect of spin polarization found that while including these degrees of freedom doesn't change phase stability for either LDA or GGA, it has appreciable effect on calculated absolute total energies. All reported calculations, other than those performed with LDA alone, were calculated to admit spin-polarization.

### A. Structural relaxation and lattice parameter calculations

Experimental data regarding lithium sites from Ref.22 which we refer to as "set B" were used as parent crystal structures for the $\gamma_a$ and $\gamma_b$ phases. Different lithium configurations were obtained by randomly removing lithium from fully filled lithium sites. Geometry optimization predicts relatively large structural relaxation for most configurations in the $\gamma_a$ phase. In some configurations lithium ions slip into positions that cannot be identified unambiguously as either a tetrahedral or octahedral site in set B for $\gamma_a$. It has been widely pointed out [21,22,43] that the designation of a site as octahedral or tetrahedral for lithium is rather ambiguous in this structure since there exist one octahedral site and two tetrahedral sites between two bond sharing octahedra. Although it is impossible for these lithium sites to be filled simultaneously due to strong lithium-lithium ion repulsion in the

$\gamma_a$ phase, residence of lithium in these sites often results from the relaxation of isolated lithium atoms in the structure. In fact, relaxation of structures in the $\gamma_a$ phase indicates that lithium ions can dwell in all lithium sites available to either the $\gamma_a$ or $\gamma_b$ phase due to the almost one-to-one mapping between these two phases. In contrast, for most configurations in the $\gamma_b$ phase initial input structures were maintained during relaxation. It is rarely found that lithium ions in this high lithiation phase relax to the lithium sites associated with the $\gamma_a$ phase. Some lithium configurations in the two-phase range, from $Li_{2.5\sim3}V_3O_8$ to $Li_4V_3O_8$, were also calculated. Large relaxations are expected in this range because the periodic boundary conditions are constraining the material's tendency to phase separate.

Lattice parameters obtained from DFT calculation using different exchange-correlation (XC) approximations are often compared with experiment under the assumption that accurate calculation of the lattice parameter will also indicate better agreement of other properties. It is usually found that GGA tends to overestimate bond length and underestimate binding energy while LDA does the opposite, when compared with experimental measurements.[44] Although the LDA/GGA rule is true for $LiV_3O_8$ system(shown in Tables I and II), the experimental data reveals non-negligible discrepancies, e.g. the experimentally measured volume of $Li_4V_3O_8$ varies by 5%. Measurements of lattice parameters for the lower lithium concentrations $Li_{1.06}V_3O_8$ (Ref.62) and $Li_{1.1}V_3O_8$(Ref.22) yield almost the same result indicating these data are perhaps more reliable. For this reason the low lithium concentration data[22] is chosen for the purpose of detailed quantitative comparison. Because realization of small fractional lithium concentrations requires a large simulation box, in practice only the lowest energy states for $Li_1V_3O_8$ and $Li_4V_3O_8$ are considered. For LDA, errors of lattice parameters $a, b, c, \beta$ and volume with respect to the experimental range vary from 0.6% to 5%; for GGA the range varies from 0.8% to 5.2%. It is found that the choice of exchange correlation functional doesn't affect the resulting lattice structure homogeneously, i.e. local bond lengths are not simultaneous shortened or lengthened. Absolute deviation of bond lengths from experiment for GGA and LDA are similar for V-O octahedra. In summary, overall agreement of bond lengths and lattice parameters between experiment and DFT calculation with either LDA and GGA are reasonable, with neither exhibiting a significant advantage over the other.

## B. Cluster expansion and phase stability

In order to study the phase stability of this compound we used a cluster expansion (CE) methodology to extrapolate from our DFT results the energies of the large number of structural states available to this compound over the composition range $0 \leq x \leq 4$. Traditional cluster expansion only approximates configuration energy well for systems with small structural relaxation. But for many crystalline materials, atomic lattice mismatch gives rise to non-negligible energy contributions. Many attempts have been made to incorporate a strain energy term into cluster expansion formalism including the mixed basis cluster expansion,[45] 'Kanzaki force' methodologies[46,47] and 'hybrid cluster expansion'.[48] In this paper we only employ a relatively simple cluster expansion formalism in which we can express the energy of the configuration state $\sigma$ by adding a single term to Eq. (1) resulting in the expression:

$$E(\sigma) = \Omega \cdot c(1-c) + J_0 + \sum_i J_i S_i(\sigma) + \sum_{j<i} J_{ij} S_i(\sigma) S_j(\sigma) + \cdots$$

(2)

where $\Omega$ represents volume deformation energy[49] and $c$ is the lithium concentration.

The experimentally derived crystal structure[22] was taken as the ideal lattice for developing this CE. As discussed in section III, large structural relaxation in the $\gamma_a$ phase enables lithium ions to reside in multiple sites that belong to the $\gamma_b$ phase. In order to apply the CE methodology to the $\gamma_a$ phase, we can set a Li-O bond length cutoff to exclude configurations with large Li-O bond lengths. In general, a more restrictive cutoff generates a smaller-error cluster expansion for the training set, but this approach would require us to discard many structures and information from a significant portion of configuration space would be lost. Another approach is to incorporate all possible lithium sites into the $\gamma_a$ phase. In particular, adding the four octahedral sites ($Li_4^\beta$, $Li_5^\beta$ multiplied by 2 due to symmetry) in the $\gamma_b$ phase was enough to capture most observed lithium displacements. Consequently the lattice model, the ideal structure upon which we develop the cluster expansion, has 12 and 10 lithium sites for the $\gamma_a$ and $\gamma_b$ phase respectively.

To create the CE we calculated the formation energies of 238 symmetrically distinct lithium configurations that were generated by randomly occupying ideal lithium for the $\gamma_a$ and $\gamma_b$ structures. Formation energies were calculated taking the total energies of $Li_1V_3O_8$, with only octahedrally coordinated lithium, and $Li_5V_3O_8$, with a fully lithiated structure, as references. So our formation energy is defined as $f$, where

$$f(Li_xV_3O_8) = E(Li_xV_3O_8) - \frac{5-x}{4}E(LiV_3O_8) - \frac{x-1}{4}E(Li_5V_3O_8)$$

(3)

Of all the configurations calculated, 136 were from $\gamma_a$ and 102 were from $\gamma_b$. Most of the configurations consisted of one primitive cell and could be described by $Li_n(V_3O_8)_2$, where n is number of lithium atoms in one primitive cell. Also 37 configurations out of 136 $\gamma_a$ structures and 21 of 102 $\gamma_b$ structures had cell sizes of 2×1×1. Because experimentally $Li_{1+x}V_3O_8$ cannot be completely delithiated, the octahedrally coordinated site $Li_1^\alpha$ is always filled for generating the initial unrelaxed structures in the $\gamma_a$ phase, and the lowest lithium composition considered in this study is $Li_1V_3O_8$.

Following standard practice in the literature cross-validation(CV) score was employed to evaluate the goodness of the cluster expansion[33]. The lower the CV score, the lower the prediction error provided by the cluster expansion. Inclusion of the volume deformation term reduced the CV score of the $\gamma_a$ phase by up to 40meV but did not appreciably lower that of high lithiation phase. This can be rationalized by the fact that configurations of the $\gamma_a$ phase experience much larger relaxations than that of the $\gamma_b$ phase, and larger relaxations give rise to larger volumetric deformations. It could be argued that since ideal lithium sites in the $\gamma_b$ phase are just a subset of that of the $\gamma_a$ phase, only one set of ideal lithium sites is needed in the cluster expansion. However, since the two phases have sigma values with a sizable difference, unique cluster expansion of two phases results in much higher CV scores.    These separate cluster expansions for the two phases are within reasonable error tolerance as shown in Table III.

DFT formation energies calculated with LDA and GGA are shown in Fig.2. At each concentration the lowest energy state was found by cluster expansion and a low energy state searching method mentioned in section II.    By comparing these two sets of formation energies, we find that GGA and LDA yield approximately the same formation energies for the $\gamma_b$ phase, e.g. the lowest formation energies at $Li_8(V_3O_8)_2$ are -0.366eV for LDA and -0.382eV for GGA. But for the $\gamma_a$ phase, GGA results in much lower values of formation energy than LDA, e.g. the lowest formation energies at $Li_4(V_3O_8)_2$ are -1.317eV for GGA and -0.969eV for LDA. It is surprising that the two phases shift relative to each other while the relative stability of the states in each phase remain largely unaffected. In experiment the two phase process initiates at around Li$_{2.5\sim3}$V$_3$O$_8$ and ends at Li$_4$V$_3$O$_8$. However, neither LDA nor GGA show two phase coexistence in the range that matches experiment. GGA indicates a two phase coexistence between Li$_{2.5}$V$_3$O$_8$ and Li$_5$V$_3$O$_8$, while the two phase region is bounded by Li$_2$V$_3$O$_8$ and Li$_4$V$_3$O$_8$ for LDA in agreement with previously reported LDA calculations[34]. In summary, both LDA and GGA predict incorrect two-phase coexistence behavior.

## C. Failure of LDA and GGA

Failure of LDA/GGA to predict correct two-phase behavior may be due to the existence of low energy states between Li$_2$V$_3$O$_8$ and Li$_4$V$_3$O$_8$ that are somehow not found by the cluster expansion. Since the cluster expansion of $\gamma_b$ phase is very accurate and the lithium configuration of lowest energy state for Li$_4$V$_3$O$_8$ here is the same as in previous work,[34] we believe the configuration of Li$_4$V$_3$O$_8$ we find is the lowest energy state and should be a ground state. While it might be possible that unidentified lower energy states exist between x=2 and x=3 such that the LDA calculation gives correct two-phase behavior, it seems unlikely that Li$_4$V$_3$O$_8$ could ever be a ground state for GGA.

A second, more plausible reason is that inappropriate treatment of strong correlations of d-electrons in vanadium ions leads to incorrect phase stability as was hinted at in the previously cited computational work.[50] To test this hypothesis, we examine the electronic structures calculated with LDA and GGA. This material is a known semiconductor[24] over the entire concentration range and resistivity increases dramatically upon lithiation.[11] However, we find that all states calculated via LDA or GGA with vanadium ion's nominal valence smaller than +5 are metallic, and only the state LiV$_3$O$_8$, in which vanadium has a nominal valence of +5, is a semiconductor. This strongly indicates that the expected lithium-host reaction where the electron contribution from the incoming lithium localizes on a single vanadium ion,[51] as described by

$$Li^+ + e^- + Li[V^{+5}]_3O_8 \rightarrow Li[V^{+5}]_2[V^{+4}]O_8 + Li^+$$

(4)

never occurs in the DFT calculations. Instead, the extra electron is delocalized in the GGA or LDA calculation. In the next section we resort to the DFT+U method to remediate this issue.

## V. DFT+U APPROACH

We performed LDA+U and GGA+U calculations on the lowest energy states at different lithium concentrations assuming that the lowest energy states are the same for DFT and DFT+U. Correct two phase behavior is recovered, and we are able to calibrate the U value by comparing these results with experimental data. In principle, the U value varies as the local structure around the vanadium ion changes and may be different for different vanadium oxidation states. However, it is only meaningful to compare energies calculated from DFT+U with the same U, and good agreement between voltages calculated using this approximation and from experiment has been reported.[52] In this study the same U was applied to all the configurations under consideration.

The need for the DFT+U formalism arises due to the emergence of strong electron correlation effects in transition metal oxide (TMO) materials, and particularly in vanadium oxide. These effects have received a significant amount of attention in the scientific literature. The material $V_2O_3$ in its paramagnetic state has long been known to undergo a pressure driven metal-insulator transition[53]. This phenomena was reproduced in dynamical mean field theory calculations with a transition observed at a Coulomb interaction U=5eV.[54] Ab initio studies of oxygen vacancies and lithium intercalation in $V_2O_5$ using the DFT+U approach have been shown to be consistent with experimental data when U=4.0eV.[55] In another systematic study of the oxidation energy of transition metals, reaction enthalpies within the vanadium oxide system (VO, $VO_2$, $V_2O_5$) were calculated by DFT+U at various U values and a range from U=3.0eV to 3.3eV were comparable with experiment.[56] It is found that U values lie in a narrow range independent of oxidation state,[55] and this value is transferable within the same system.[56] These phenomena are not unique to the vanadium oxide system. Charge ordering and Jahn-Teller distortion also occur in spinel $LiMnO_2$[57,58] and $NaCoO_2$.[59] While these effects are not observed in DFT calculations they can be reproduced within the DFT+U formalism. There are notable exceptions to the degree of improvement offered by this approach. Overall performance of GGA was shown to be better than GGA+U for $NaCoO_2$[59] since switching from GGA to GGA+U involves the disappearance and emergence of ground states and the total number of ground states in DFT+U is reduced.

Because DFT+U predicts $LiV_3O_8$ to be a semiconductor rather than a metal, a smaller k-point density mesh 4×6×2 is required to obtain energies of the same accuracy. In this work, we use the simplified approach of Dudarev[28] as implemented in VASP. We will simply use U to denote effective term U-J.

Table I,II list lattice constants of $Li_1V_3O_8$ and $Li_4V_3O_8$ calculated from DFT+U, respectively. Loschen, et al.[60] in their DFT+U study of cerium oxides found that lattice parameter increases steadily with growing U. We also found general volume expansion of $Li_{1+x}V_3O_8$ by increasing U both for GGA and LDA. However, $Li_4V_3O_8$

showed much more significant expansion than $Li_1V_3O_8$. For $Li_1V_3O_8$, change of lattice parameters and V-O, Li-O bond lengths are negligible.

Fig.3 shows our results for both LDA+U and GGA+U calculations with various values of U. It can be seen that, for both LDA and GGA, as the value of U increases the depth of the convex energy valley is greatly reduced. This effect on formation energies is similarly to that seen in work by Zhou et al. [61] where they found initially negative formation energies of $Li_xFePO_4$ increase with U and become positive at U=2.5~3.5eV. Also, in their work, formation energy converged with respect to U, however in our study of $Li_{1+x}V_3O_8$, formation energy diverges when U =3 for LDA while it seems to converge at the same value of U for GGA. Occurrence of phase coexistence between $Li_{2.5}V_3O_8$ and $Li_4V_3O_8$ was seen with U>=2 for LDA and U>=1 for GGA. Thus it is suspected that with an appropriate value of U, the correct phase stability and thermodynamics can be recovered.

## A. Electronic structure

One of the shortcomings of both the LDA and GGA calculations performed in the previous section was the fact that, although these materials are known semiconductors over the entire composition range, the calculations predict them to be metallic except when maximally delithiated. Fig.4 shows the density of states for this maximally delithiated state $Li_1V_3O_8$ calculated using both GGA and GGA+U. Hybridization of O p band and V d band results in a filled lower bonding band and an unoccupied antibonding band. The shape of the DOS calculated from GGA+U doesn't change with increasing U except that the band gap increases from 0.9eV to 1.3eV. Upon lithiation, the band structure of $Li_2V_3O_8$ doesn't change from $Li_1V_3O_8$ within the GGA calculation, and one electron per lithium is donated to fill the vanadium d orbital in the lower end of conduction band. As a result the Fermi level is shifted up into conduction band as shown in Fig.4. Further lithiation gives the same result except for the magnitude of Fermi level shift. Integration of the newly filled total DOS is equal to the number of lithium ions incorporated. Inspection of the partial DOS for vanadium ions shows that all vanadium ions share the contribution from the 3d electrons nearly equally.

For both LDA and GGA as the value of U increases the density of states around the Fermi level decreases to zero(Fig.4), indicating a transition from a metal to a semiconductor. This split of the conduction band is the result of d electron localization due to the energy penalty determined by the value of U. For the case of $Li_2V_3O_8$(see Fig.5), the DOS splits only in the majority spin channel of V(3) while the minority spin channel is kept intact for all vanadium ions. As a result, the lower part of the DOS of V(3) is filled, which means that these ions are in valence states $V^{4+}$ while other ions are in valence state $V^{5+}$. (Although hybridization of Op-Vd band makes vanadium ionization less than the nominal ionization, we still designate the ionization state to be $V^{5+}$ for $LiV_3O_8$ as a reference valence state.) For the purpose of comparison, the DOS from LDA calculation and LDA+U are also shown in Fig.4. GGA and LDA essentially share the same shape DOS and the same trend with increasing U, but GGA seems to be more susceptible to the energy penalty imposed by U.

Since this material is empirically observed to be a semiconductor over the entire concentration range, we need to choose a large enough value of U such that all structures become semiconducting. The experimental measurement of $Li_1V_3O_8$ single crystal[62] indicates a small bandgap of approximately 0.1eV, which is smaller than the value predicted by either DFT or DFT+U calculations for $Li_1V_3O_8$ where the band gap is approximately 1.0eV. However, for higher values of lithiation, $Li_{1+x}V_3O_8$, we can technically open a small gap within the 3d band with appropriate U for each configuration. In practice, we found that the U value required to achieve this small bandgap in different configurations varies considerably. As mentioned above, however, for purpose of energy comparison we must use the same U for all configurations. For GGA most of the lowest energy states become semiconducting with U=3eV. On the other hand for LDA phase behavior significantly deviates from experimental observations with U=3eV and most lowest energy states are still metallic.

We can further test the validity of these calculations by investigating whether these calculations predict preferential reduction associated with small polaron transport. M Onoda et al[62] found highly anisotropic resistivity in $Li_{1+x}V_3O_8$ and significant difference in energy gaps calculated from temperature dependent resistivity and thermoelectric power measurements, which suggests small polaron motion. This small polaron transport is commonly found in many other semiconducting compounds, e.g. electron polaron in $Li_xFePO_4$ and hole polaron in $Li_{1-x}FePO_4$[63]. The electron donated from lithium intercalates into toptactic compound to form an electron polaron which hops from one transition metal ion to another. In particular, M Onoda[62] found small polorans mainly exist on the V(2) and V(3) sites due to preferential reduction. In another study, Florent Boucher[64] also showed a preferential reduction sequence V(3)>V(2)>V(1) upon lithiation from XPS data.

In order to see if the GGA+U method also yields the same preferential reduction, we evaluate vanadium valance states for $Li_2V_3O_8$ and $Li_5V_3O_8$(see Fig.5), which have 2 and 8 donated electrons respectively. Since vanadium ions are in the same valence state in $Li_1V_3O_8$ and $Li_4V_3O_8$, e.g. $V^{5+}$ for $Li_2V_3O_8$ and $V^{4+}$ $Li_5V_3O_8$, valence states in

Li$_2$V$_3$O$_8$ and Li$_5$V$_3$O$_8$ will show which site the electron preferentially occupies upon incorporation of one more lithium. These structures are chosen such that they are the lowest energy states at their respective compositions. As discussed above, in Li$_2$V$_3$O$_8$ two V(3) are in valance state V$^{4+}$ while V(1) and V(2) are still in valence state V$^{5+}$. Similarly in Li$_5$V$_3$O$_8$ where all d-bands of V sites involve gap opening, V(1) and V(2) each receive one electron and V(3) receives two electrons, which reduce them to V$^{4+}$,V$^{4+}$ and V$^{3+}$ respectively. These results agree very well with the preferential reduction in V found in experiments, indicating that the GGA+U calculation provides reliable result beyond the calculation of formation energies.

## B. Voltage

The relative voltage of a cathode material in a lithium ion battery is one of its most important characteristics for determining its performance. It is defined as[32]:

$$V = \frac{\mu_{cathode}^{Li} - \mu_{anode}^{Li}}{F}$$

(5)

where $\mu_{cathode}^{Li}$, $\mu_{anode}^{Li}$ are the chemical potentials of lithium in the cathode and anode respectively and $F$ is Faraday's constant. Calculation of $\mu_{cathode}^{Li}$ requires free energy information which is computational costly to obtain, thus only average voltage is considered here. The average voltage between two lithium compositions Li$_{x1}$V$_3$O$_8$ and Li$_{x2}$V$_3$O$_8$ can be calculated using following formula[52]:

$$V = \frac{E(Li_{x_2}V_3O_8) - E(Li_{x_1}V_3O_8) - (x_2 - x_1)E(Li)}{(x_2 - x_1)F}$$

(6)

where E(Li$_x$V$_3$O$_8$) stands for total energy of Li$_x$V$_3$O$_8$. In our calculation, bcc lithium metal is taken as the anode material[65].

As shown in Fig.6, calculated voltages for both LDA and GGA were found to be lower than the measured experimental values. With increasing values of effective U, calculated voltages tend to more closely approach the experimental measurements. For GGA, U=3eV was found to have calculated voltages that closely match the experimental curve except at very low lithium concentrations where the calculated voltage is much lower and has no tendency to increase farther with U. For LDA, U=2eV gives calculated voltages that only roughly overlap with experiment data. However, LDA+U captures the rapid voltage decrease with increasing lithiation at low lithium concentration.

## C. Phase stability

In the previous two sections we have shown that GGA+U calculations with U=3eV yield results that are generally in good agreement with experiment. Using this formalism we calculated the formation energy for approximately 90 configurations as shown in Fig.7. These calculations confirmed that all the ground states using GGA+U are identical to the lowest energy states calculated with GGA or LDA. We did not attempt to cluster expand the formation energy using data calculated in GGA+U, since this would require taking into account Columbic interactions between ions in different valence states.[59]    For the $\gamma_a$ phase, a large set of configurations between compositions $Li_{1.125}V_3O_8$ and $Li_{1.75}V_3O_8$ were calculated and the lowest energy states were found to fall on the convex hull indicating solid solution behavior for a concentration range from $Li_1$ to $Li_5$. For the $\gamma_b$   phase, $Li_4V_3O_8$, $Li_{4.25}V_3O_8$, $Li_{4.5}V_3O_8$, $Li_{4.75}V_3O_8$ and $Li_5V_3O_8$   were found to lie almost precisely on a straight line. Formation energies of $Li_{4.25}V_3O_8$, $Li_{4.75}V_3O_8$ are above convex line, which connects $Li_4V_3O_8$, $Li_{4.5}V_3O_8$ and $Li_5V_3O_8$, by 0.04meV and 2meV respectively.    These energy differences are about the numerical error and $\gamma_b$ phase is believed to be a solution phase at finite temperature. X-ray diffraction simulations of structures of Li$_{1+x}$V$_3$O$_8$ at x=0, 0.5, 1,1.5, 3,4 are shown in Fig.8 . The primary experimental X-ray diffraction evidence of Li$_{1+x}$V$_3$O$_8$ undergoing a phase transition is a shift of the (100) peak, which is mainly due to an increase of lattice constant *a* from that of the $\gamma_a$ phase to that of the $\gamma_b$ phase. Fig.8 clearly shows a shift of the (100) peak between the simulated X-ray diffraction data from Li$_{2.5}$V$_3$O$_8$ and Li$_4$V$_3$O$_8$.

Some of the ground state structures are shown in Fig.9. We use the notation (1) and (2) to denote two symmetrically equivalent sites in the two-molecular unit cells. We find that these ground state structures generally follow the rule of maximal separation of lithium ions. Obtaining the ground state of Li$_{1.5}$V$_3$O$_8$ from Li$_1$V$_3$O$_8$ by placing another lithium on the $Li_2^\alpha$ (2)(or $Li_2^\alpha$ (1) )site has been correctly predicted by many others [21,50] based on the minimum lithium-lithium repulsion argument. The ground state of Li$_2$V$_3$O$_8$ cannot be obtained by putting lithium on the $Li_2^\alpha$ (1)(or $Li_2^\alpha$ (2) ) site due to the short $Li_2^\alpha$ (1)- $Li_2^\alpha$ (2) distance. The incoming lithium must reside in (1) if the $Li_2^\alpha$ site in (2) is filled, and its optimal location is at same fractional coordinate in the b direction as $Li_1^\alpha$ to avoid strong repulsion from filled $Li_2^\alpha$ site.  We found the ground state of Li$_2$V$_3$O$_8$ consists of two pairs of lithium ions: ($Li_a^\alpha$ (1), $Li_b^\alpha$ (1))  at b=1/4 and unchanged ($Li_1^\alpha$ (2), $Li_2^\alpha$(2)). The ground state of Li$_{2.5}$V$_3$O$_8$ also follows this rule: ($Li_2^\alpha$ (1), $Li_a^\alpha$(1), $Li_b^\alpha$(1)) at b=1/4 and ($Li_a^\alpha$ (2), $Li_b^\alpha$ (2)) at b=3/4. These almost uniformly distributed and alternating groups of lithium ions require minimal interaction between lithium ions, and for this reason are more likely to be ground states. Based on the lithium configurations of the lowest energy states in the $\gamma_b$ phase before onset of the two-phase process, we propose a more plausible lithiation sequence than previously proposed.[22] From Li$_1$V$_3$O$_8$ to Li$_{1.5}$V$_3$O$_8$, one tetrahedral site $Li_2^\alpha$ (2) (or $Li_2^\alpha$(1)) is filled. From $Li_{1.5}V_3O_8$ to $Li_2V_3O_8$, rather than filling up another $Li_2^\alpha$ (1) (or $Li_2^\alpha$ (2)) site, the $Li_1^\alpha$ (1), (or $Li_1^\alpha$ (2),) is driven by another incoming lithium to jointly occupy two octahedral sites $Li_a^\alpha$ (1) (or $Li_a^\alpha$ (2)) and $Li_b^\alpha$ (1) (or $Li_b^\alpha$ (2)). This process is kinetically convenient because $Li_1^\alpha$ and $Li_a^\alpha/Li_b^\alpha$ are in two face-sharing octahedra. These two octahedrally coordinated lithium ions lean toward each other distorting surrounding oxygen-ligands such that they can be treated as if they are in tetrahedral sites when longer oxygen bonds are ignored. From $Li_2V_3O_8$ to $Li_{2.5}V_3O_8$, the $Li_1^\alpha$ (2) (or $Li_1^\alpha$ (1)) undergoes the same process that the previous $Li_1^\alpha$(1) (or $Li_1^\alpha$ (1)) underwent. Upon further lithiation a phase transition takes place between Li2.5 –Li4. Incoming lithium ions hop into the octahedral site directly and, together with the distorted octahedral sites $Li_a^\alpha, Li_b^\alpha$ and tetrahedral site $Li_2^\alpha$, adopt a less distorted local structure comprised of $Li_1^\beta, Li_2^\beta, Li_3^\beta$ and $Li_4^\beta$. This process also involves a sudden shortening of lattice constant a due to dramatic increase of lithium concentration and a decrease of the V1-O7 bond length to form a regular $VO_6$ octahedron.

## VI. LITHIUM DIFFUSION

Intercalation compounds can be categorized by the dimensionality of the lithium diffusion process by which charging and discharging proceeds. Spinels of type $LiTiS_2$ and $LiMnO_2$ exhibit three dimensional lithium diffusion paths.[70,71] Layered transition metal oxides typically undergo a two-dimensional lithium diffusion process, e.g. $LiCoO_2$. It is found that the migration energy decreases with increasing lithium concentration in layered $LiCoO_2$ and spinel $LiTiO_2$.[32,72] The lithium diffusion mechanism in $LiFeO_4$ was first explored by ab inito calculation and was discovered to be one-dimensional, which has important consequences for improving the performance of batteries that utilize this electrode composition.[73] In order to study lithium diffusion in $Li_{1+x}V_3O_8$ we calculated the migration energy of lithium hopping in different configurations via the elastic band methodology as implemented in VASP. For all the calculations we used a 2×3×1 supercell and a single gamma point Brillouin zone integration. Although different opinions exist regarding the importance of electron-correlations on migration barriers,[71,74] we utilized DFT+U calculations to ensure possible electron-correlation effects are included.

As discussed previously, the $Li_{1+x}V_3O_8$ crystal structure can be visualized as a stacking of zigzagged $V_3O_8$ layers, thus our expectation would be that lithium diffusion should be constrained in two dimensions. Three lithium migration paths were chosen and minimum energy paths were calculated at two lithium concentrations, $Li_4V_3O_8$ and $Li_5V_3O_8$, representing the high lithiation phase. The direction vector for these three migration paths were, respectively, (-0.10,0,-0.16)( $Li_3^\beta \rightarrow Li_4^\beta$), (-0.25,-0.49,-0.10) ($Li_5^\beta \rightarrow Li_4^\beta$) and (0.57 0.01,0.01) ($Li_3^\beta \rightarrow Li_5^\beta$). These lithium migration paths are chosen such that path 1 and path 2 are on the bc plane between vanadium oxide layers. Path 1 is almost parallel to the vanadium oxide plane; path2 has non-zero components in all three directions, and path 3 is parallel to the *a* axis, across the vanadium oxide plane.

Energy landscape along migration path is given in Fig. 10. Because of the low symmetry of the $Li_{1+x}V_3O_8$ crystal structure, all of the migration paths show asymmetric energy as a function of the reaction coordinate. Thus, the migration rates across the saddle points are direction dependent. For the high lithiation phase $Li_4V_3O_8$, diffusion energy between bc planes varies from 0.08eV to 0. 62eV, while diffusion energy across vanadium oxide planes is about 1.2eV. Although diffusion energy along other migration paths are not all known, for qualitative analysis, it is reasonable to conclude that at room temperature diffusion on bc planes will dominate the diffusion process in $Li_4V_3O_8$. For a fully lithiated structure $Li_5V_3O_8$, we find diffusion anisotropy is mitigated. Diffusion energy between vanadium oxide layers range from 0.31eV to 0.87 eV while the barrier across vanadium layers is about 0.9eV. This is reasonable since fully lithiated $Li_5V_3O_8$ has a more ordered rock-salt type structure, which possesses higher symmetry, since all Li-O, V-O octahedra are identically coordinated.

Further investigation of the relation between lattice parameters and migration energy indicates a negative correlation. The lattice constant along a direction decreases from 6.2233 for $Li_4V_3O_8$ to 6.0252 for $Li_5V_3O_8$ due to the "gluing" effect of lithium between vanadium oxide layers. Path 1, which is parallel to vanadium oxide layers, exhibits a dramatic increase in migration energy in both hopping directions: from 0.08 to 0.31eV and from 0.24 to 0.87eV. It also has been shown that expansion of interlayer spacing due to incorporation of inorganic compounds such as $H_2O$ or $CO_2$ increases lithium ion mobility.[66] The center of the migration path between two octahedral sites often corresponds to a high symmetry point in the ordered crystal structure, i.e. a tetrahedral site or another octahedral site. In this case, the local minimum corresponds to a tetrahedral site, which shares a triangular face with one of the octahedra. Note that for path 3, the saddle point, which corresponds to an octahedral site between two other octahedral sites, is not a local minimum. Measurement of the volume of the octahedron containing the mobile lithium at the saddle point gives 11.64Å$^3$ for $Li_4V_3O_8$ and 12.31 Å$^3$ for $Li_5V_3O_8$, indicating that lower lithium concentration provides a more confined channel for lithium transport across vanadium oxide layers. This lithium concentration effect was manifested in a much lower migration barrier ~0.9eV in $Li_5V_3O_8$ than ~1.2eV in $Li_4V_3O_8$.

Three different migration paths, which we will again refer to as paths 1 – 3, were selected to study lithium diffusion behavior in the low lithiation phase $Li_{1.5}V_3O_8$: (-0.10,0,-0.24)( $Li_2^\alpha \rightarrow Li_a^\alpha$), (0.35,0,-0.04)( $Li_a^\alpha \rightarrow Li_b^\alpha$), (0.15,-0.45,0.12)( $Li_a^\alpha$ (2)$\rightarrow$ $Li_a^\alpha$ (1)). Paths 1 and 3 exhibit migration energies between 0.15~0.36eV, which roughly agrees with NMR spectrometry measurement of $Li_{1.1}V_3O_8$, in which high lithium mobility at room temperature was measured to exhibit an activation energy for self-diffusion of ~0.31eV. [67] Migration path 2 evinces a much larger barrier to lithium migration ranging from 0.43 to 0.61eV. Investigation of the structural environment of migration along path 2 shows that near the transition state the mobile lithium ion is sandwiched between two other lithium ions that are 3.6A from each other in the absence of relaxation. After relaxation, the distance between these two lithium ion expands to 5.2A indicating a strong repulsion between lithium ions at distances smaller than 1.8A. Since this need to pass close to other lithium ions is not observed along migration path 1 or path 3, the lithium ions around the moving lithium ion are less distorted during the transition. Vanadium oxide octahedra are more rigid than Li-O octahedra and this distortion leads to a change in lattice parameter that, in turn, affects migration energies.

## VII. CONCLUSION

Our combined DFT+CE and DFT+U calculation approach presented above has revealed some aspects of thermodynamics and kinetics which are more consistent with experiment than previous DFT calculations.[34] In particular, the full relaxation of the lattice parameter during DFT calculation of formation energies was found to be critical for finding correct ground states and low energy states in the $Li_{1-x}V_3O_8$ system. While DFT calculation results using LDA and GGA exhibit similar phase stability, they nevertheless both fail to predict a composition range for the two-phase process that is in agreement with experiment. Our investigation shows that the failure of LDA and GGA is most likely due to the strong correlation of vanadium d electrons, which can be accounted for adequately within a DFT+U framework.    The structures of the lowest energy states at different lithium compositions, which were determined by a CE searching method, are observed to remain unchanged by switching from DFT to DFT+U. A coexistence between $Li_{2.5}V_3O_8$ ( $\gamma_a$  phase) and $Li_4V_3O_8$($\gamma_b$ phase) is evident from the DFT+U calculations with an appropriate choice for the U value.    This DFT+U method is validated by comparison of the calculated densities of states with the reported semiconducting behavior of the compounds and from comparison to experimental findings regarding the preferential reduction of vanadium ions. By testing the voltage curve and the semiconducting electronic band structure at various U values, we found an optimized value of U≈3eV for GGA. However, no satisfactory U value can be found for LDA because of divergence in the formation energy at larger U values and no recovery of semiconductivity at smaller U. Analysis of the ground state structures calculated from the DFT+U method provides a key to understanding the phase transformation between $\gamma_a$ and $\gamma_b$ and a plausible lithiation sequence for this compound. Note that we make no claim that this is the only possible transition pathway, as other transition pathways are likely to occur experimentally since batteries tested in laboratory are often driven into non-equilibrium or metastable states as demonstrated in kinetic Monte Carlo simulation.[68] This is also likely one of the reasons that no consensus on onset of two-phase process for $Li_{1+x}V_3O_8$    is seen in literature.

Anisotropy of lithium diffusion is generally determined by anisotropy of lithium migration barriers along different diffusion channels in the host lattice structure, however for systems like $Li_{1+x}V_3O_8$ with strong lithium-host interactions, diffusion becomes less anisotropic as lithium concentration increases because the lattice becomes more ordered. Migration barriers were found to depend on changes of the lattice constant *c*, confinement of the diffusion channel at the saddle point and lithium-lithium repulsion around the saddle point, which show notable agreement with previous studies of lithium mobility in layered lithium transition metal oxides[69] where activation energy was reported to be strongly altered by the size of tetrahedral sites at saddle points and lithium-lithium electrostatic repulsion.

## ACKNOWLEDGEMENTS

This work was funded by the U.S. National Science Foundation Cyber Discovery and Innovation program under Grant No. CHE1027765. Computational work was performed at the JHU Homewood High-Performance Compute Cluster at the Institute for Data Intensive Engineering and Science. The authors also acknowledge valuable conversations and guidance from Prof. Anton Van der Ven.

Figure caption

FIG.1. Lithium sites available within the $\gamma_a$ and $\gamma_b$ phases. The sites available in $\gamma_a$ are shown in (a) and consist of one octahedral site $Li_1^\alpha$ and three tetrahedral sites $Li_2^\alpha, Li_a^\alpha, Li_b^\alpha$. The five sites available in $\gamma_b$ are all octahedral sites, and these are show in (b) as $Li_1^\beta, Li_2^\beta, Li_3^\beta, Li_4^\beta, Li_5^\beta$. V(1),V(2),V(3) are labeled and are the same for both the $\gamma_a$ and $\gamma_b$ phases.

FIG.2. Formation energies calculated with (a) GGA and (b) LDA. Every point represent a unique configuration. Each line connects ground states on the convex hull.

FIG.3. Formation energies of the lowest energy states calculated using DFT+U with various U values for LDA+U(a) and GGA+U(b).

FIG.4. DOS of $Li_1V_3O_8$ calculated by (a) GGA and (d) GGA+U=3eV. DOS of $Li_2V_3O_8$ calculated by (b) GGA, (e) GGA+U=3eV, (c) LDA, and (f) LDA+U=3eV. The d band in $Li_2V_3O_8$ splits at U=3eV around the Fermi level for GGA, while it hasn't opened a gap for LDA.

FIG.5. V-d partial density of states for $Li_2V_3O_8$ and $Li_5V_3O_8$ calculated by GGA+U=3eV. DOS (x1) and (x2) denote the V-d band on the two V(1) sites, (x3) and (x4) on the two V(2) sites and (x5) and (x6) on the two V(3) sites, where x denotes either the (a) $Li_2V_3O_8$ or (b) $Li_5V_3O_8$ structure.

FIG.6. Voltage curve calculated from (a) LDA+U and (b) GGA+U. The dotted line is the open circuit voltage(OCV) measured experimentally in Ref. 75.

FIG.7. Formation energies of $Li_xV_3O_8$ calculated by GGA+U=3 eV.

FIG.8. Simulated x-ray powder diffraction for (a) $Li_1V_3O_8$ (b) $Li_{1.5}V_3O_8$ (c) $Li_2V_3O_8$ (d) $Li_{2.5}V_3O_8$ (e) $Li_4V_3O_8$ (f) $Li_5V_3O_8$. Note the a dramatic shift of the (100) peak between (d) and (e).

FIG.9. Ground states of lithium composition at (a) $Li_1V_3O_8$, (b) $Li_{1.5}V_3O_8$, (c) $Li_2V_3O_8$, (d) $Li_{2.5}V_3O_8$, (e) $Li_4V_3O_8$, and (f) $Li_5V_3O_8$ .

FIG.10. Energy landscape along selected migration paths (a1)-(a3) for $Li_5V_3O_8$, (b1)-(b3) for $Li_4V_3O_8$, (c1)-(c3) for $Li_{1.5}V_3O_8$. For each composition the number represents path1 to path3 as described in the text. Moving lithium at left and right end of the graph for each migration path are (a1)(b1) ($Li_3^\beta \rightarrow Li_1^\beta$), ($Li_5^\beta \rightarrow Li_4^\beta$) ,(a2)(b2) ($Li_5^\beta \rightarrow Li_4^\beta$ ) , (a3)(b3) ($Li_3^\beta \rightarrow Li_5^\beta$) , (c1) ($Li_2^\alpha \rightarrow Li_a^\alpha$), (c2) ( $i_a^\alpha \rightarrow Li_b^\alpha$), (c3) ($Li_a^\alpha$ (2)$\rightarrow Li_a^\alpha$(1)). Migration barriers for lithium hoping from the left and from the right for each migration paths are (a1) (0.31eV,0.87eV), (a2)(0.62eV,0.11eV), (0.15eV,0.36eV), (a3)(0.87eV,0.96eV), (b1)(0.08eV,0.24eV), (b2)(0.62eV,0.35eV), (b3)(1.11eV,1.09eV), (c1)(0.34eV,0.16eV), (c2)(0.61eV,0.43eV), (c3)(0.36eV,0.20eV).

TABLE I. Lattice parameters from experiment and calculation ($Li_1V_3O_8$ )for the $\gamma_a$ phase.

TABLE II. Lattice parameters from experiment and calculation ($Li_4V_3O_8$) for the $\gamma_b$ phase.

TABLE III. Cluster expansion fitness values.

Figure 1

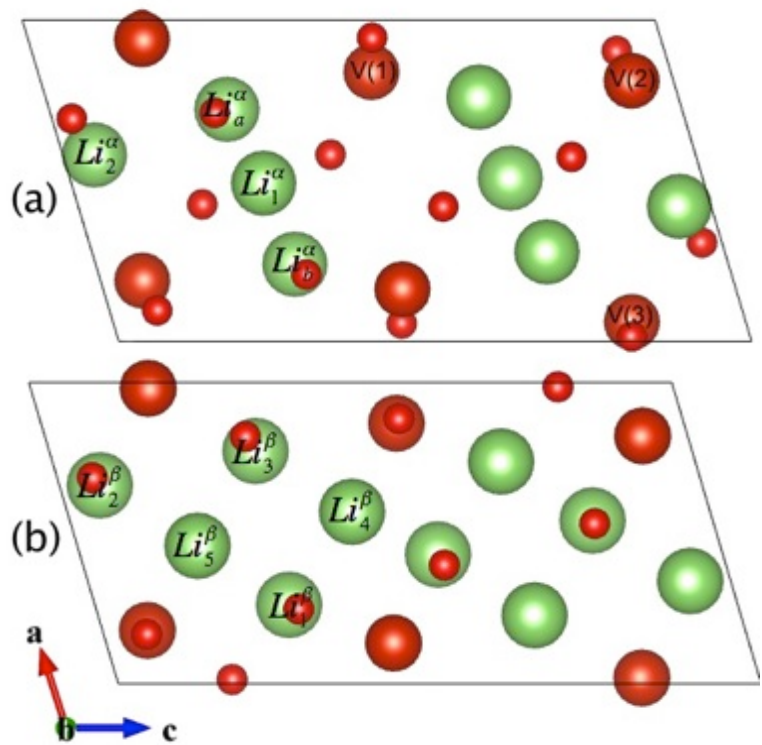

Figure 2

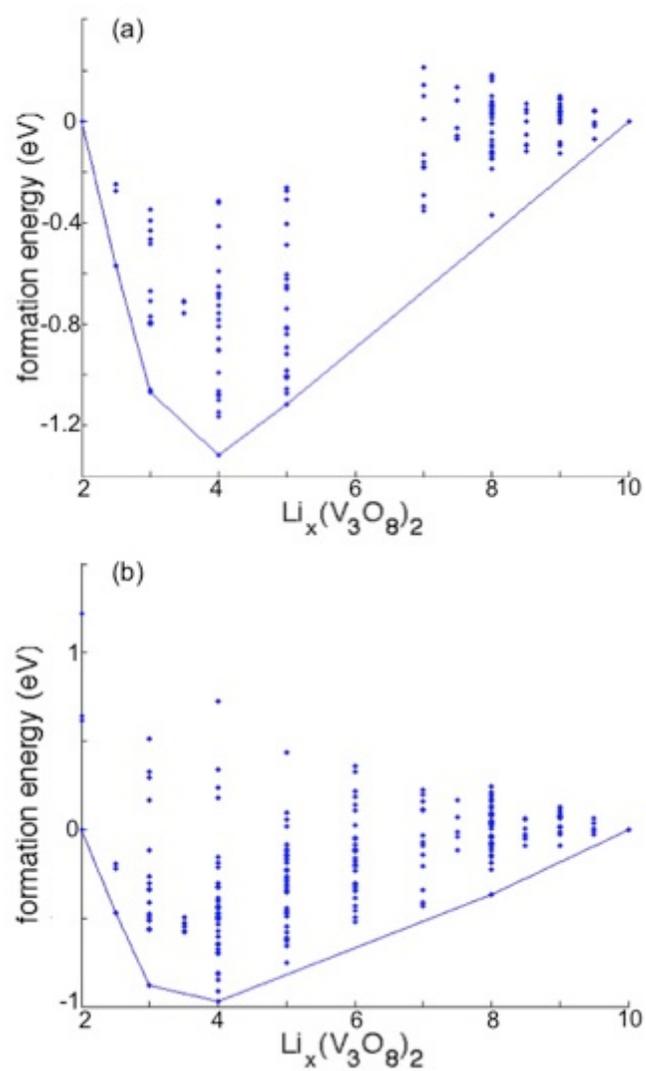

Figure 3

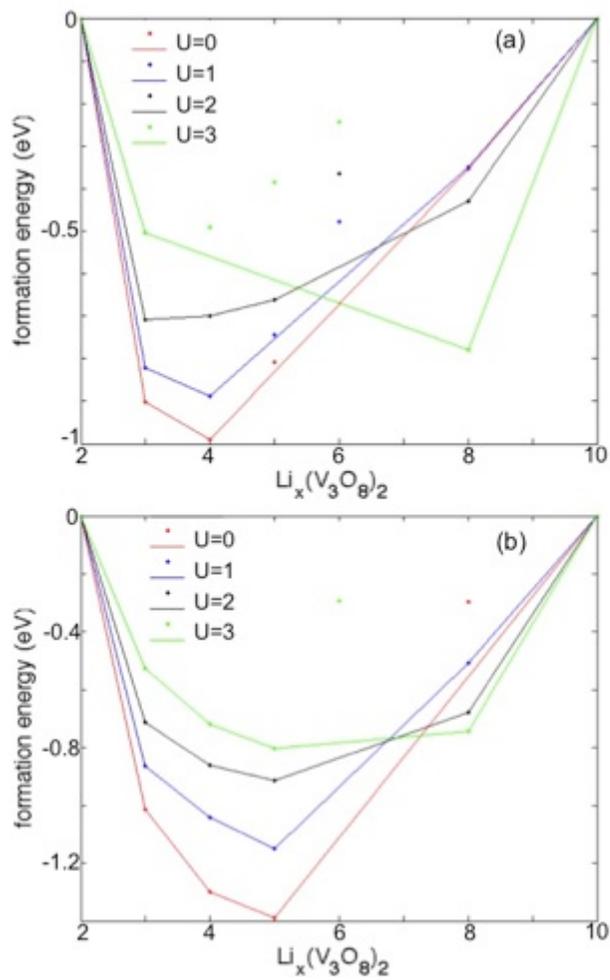

Figure 4

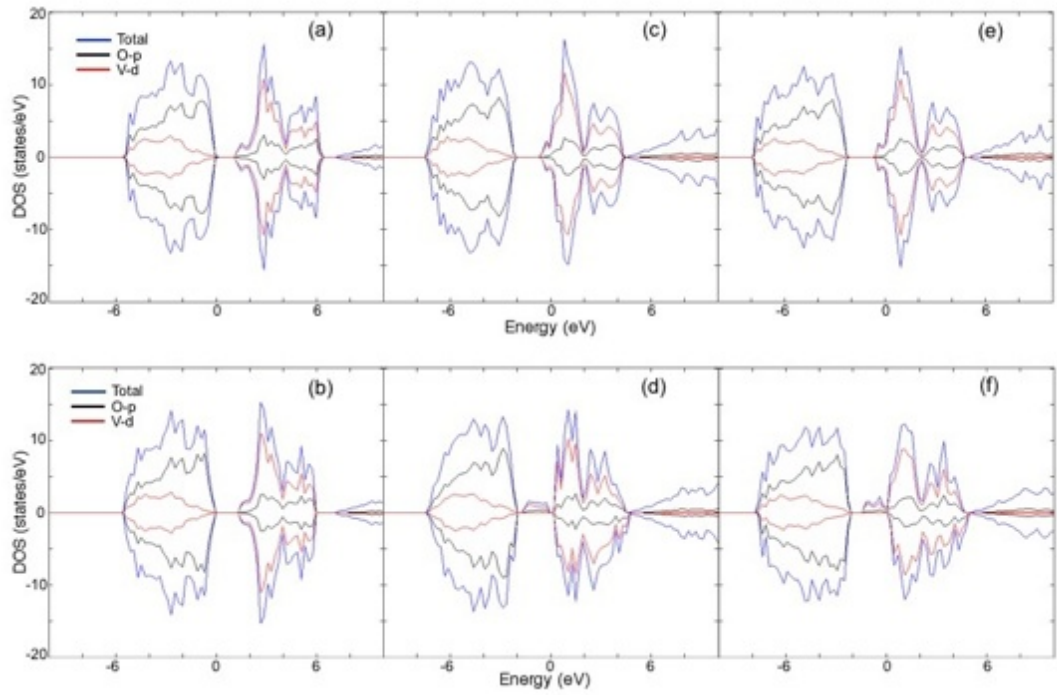

Figure 5

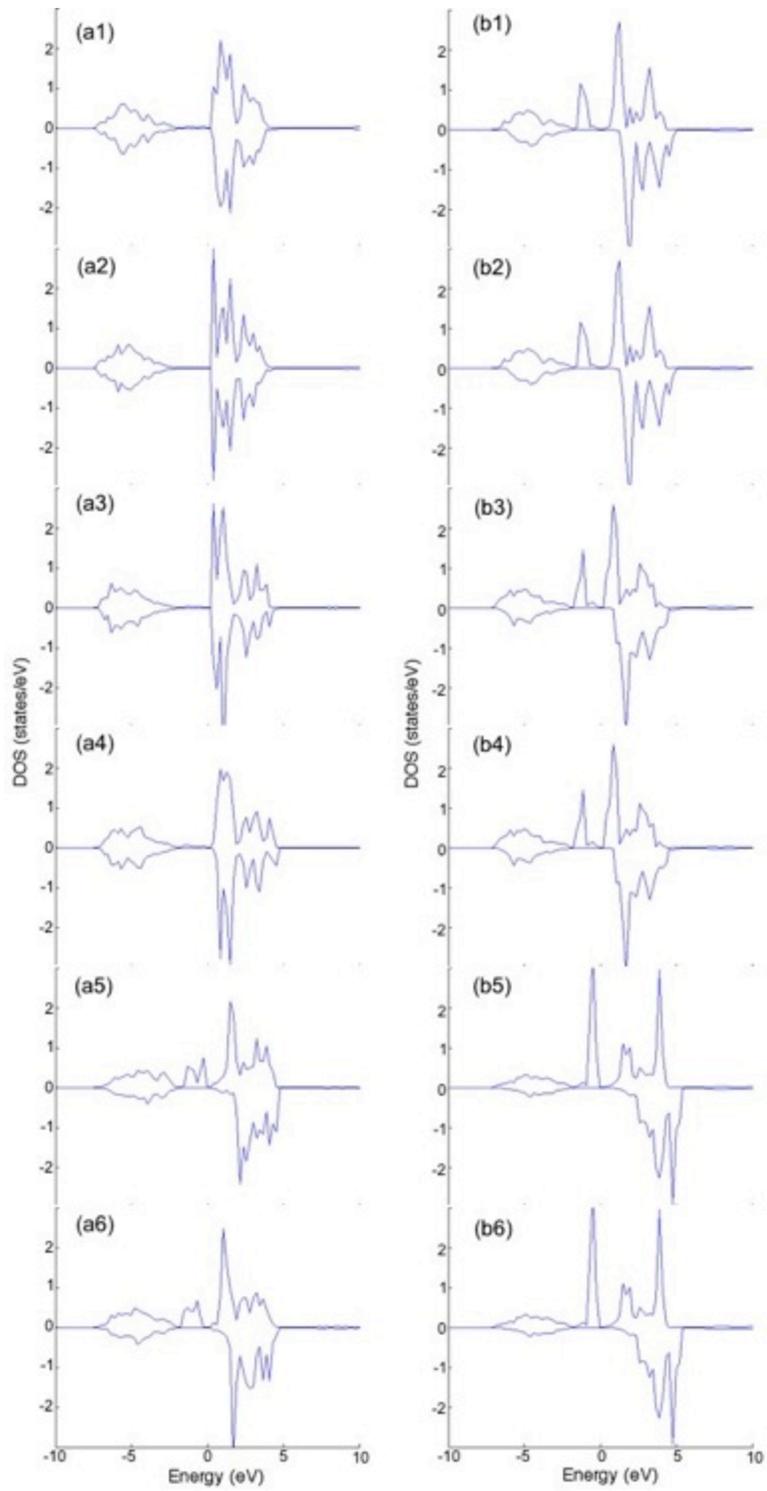

Figure 6

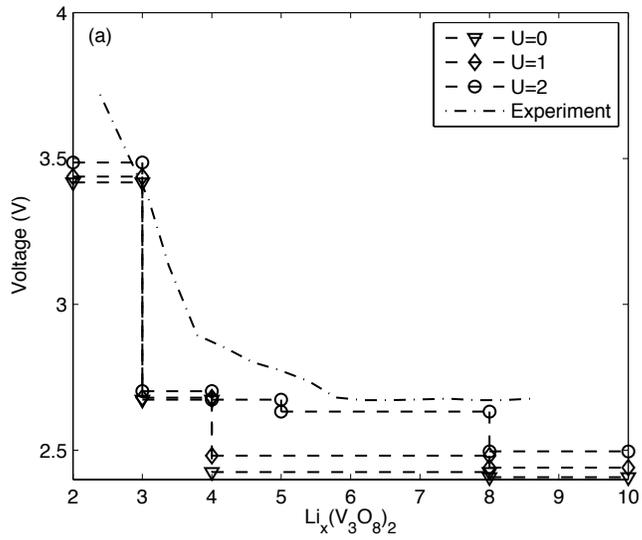

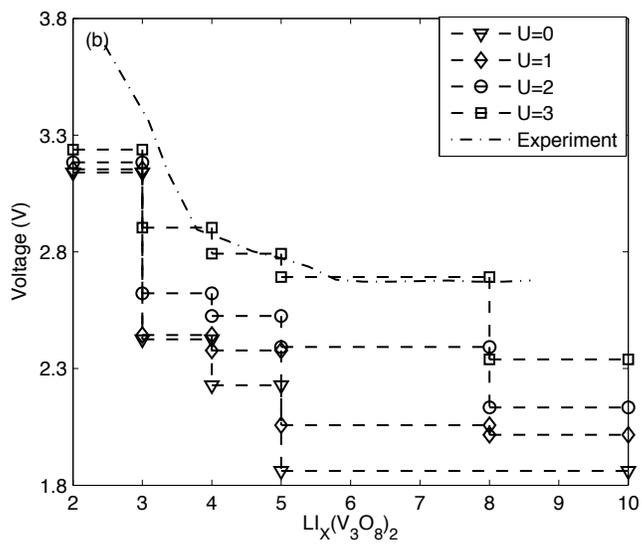

Figure 7

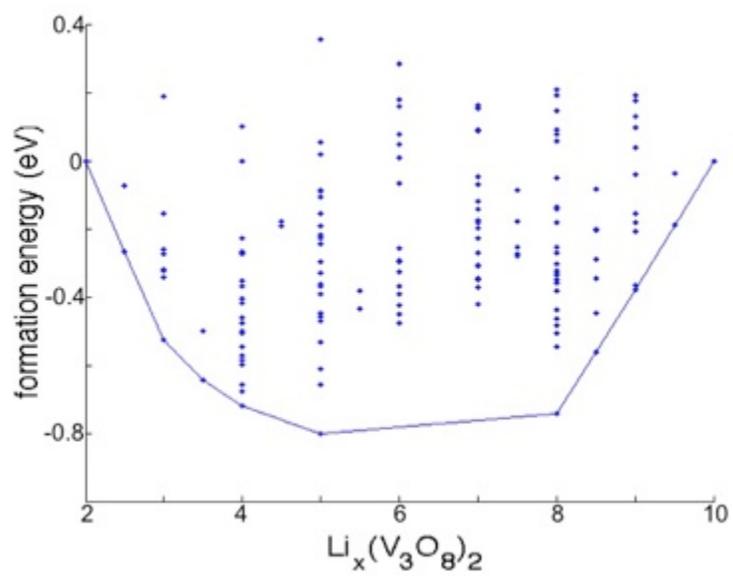

Figure 8

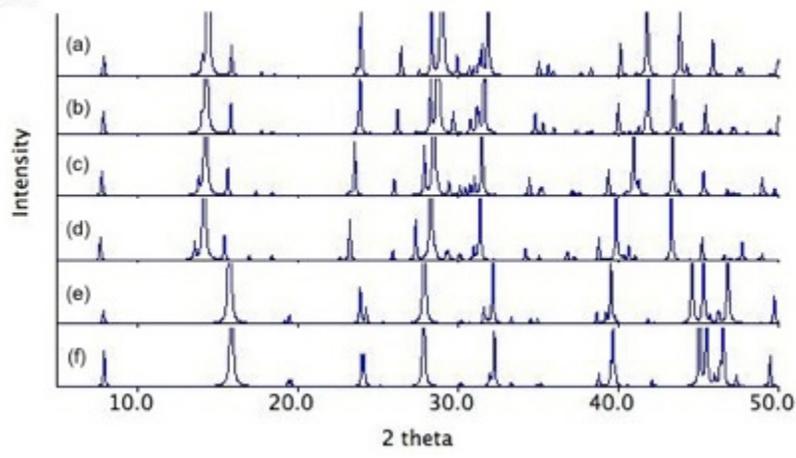

Figure 9

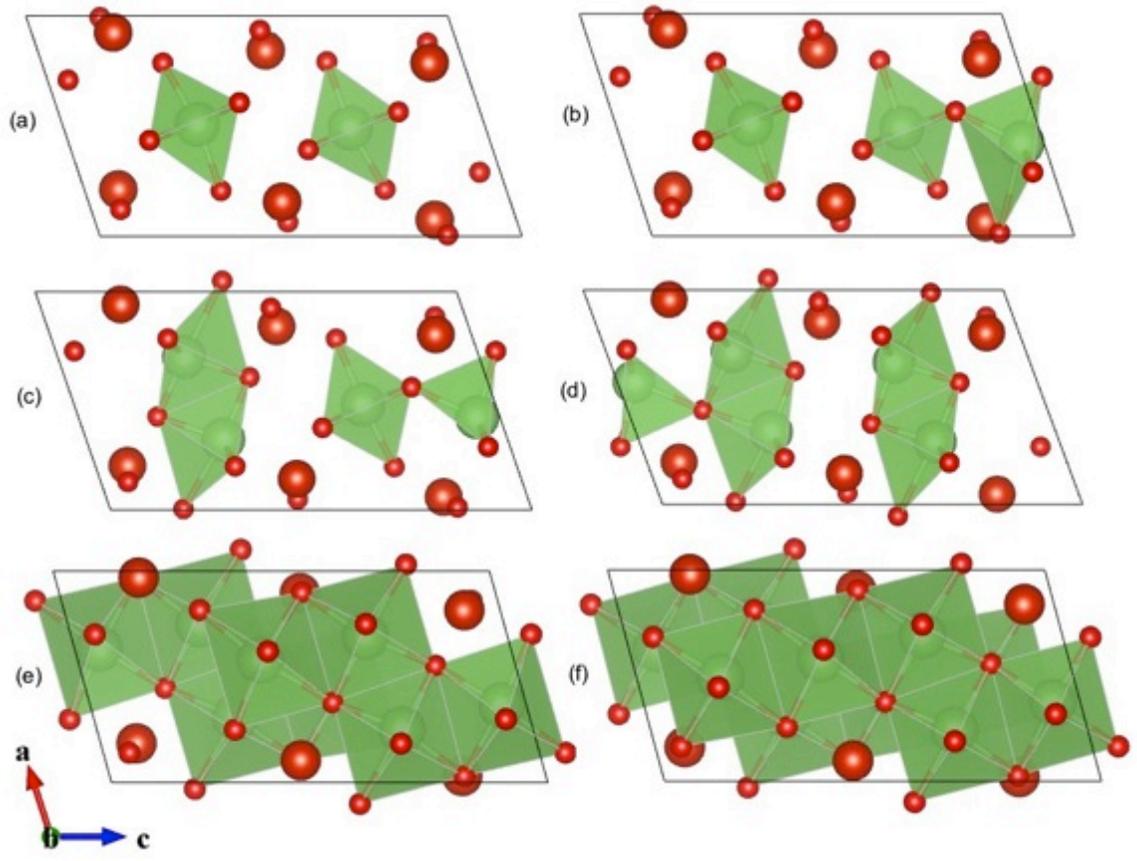

Figure 10

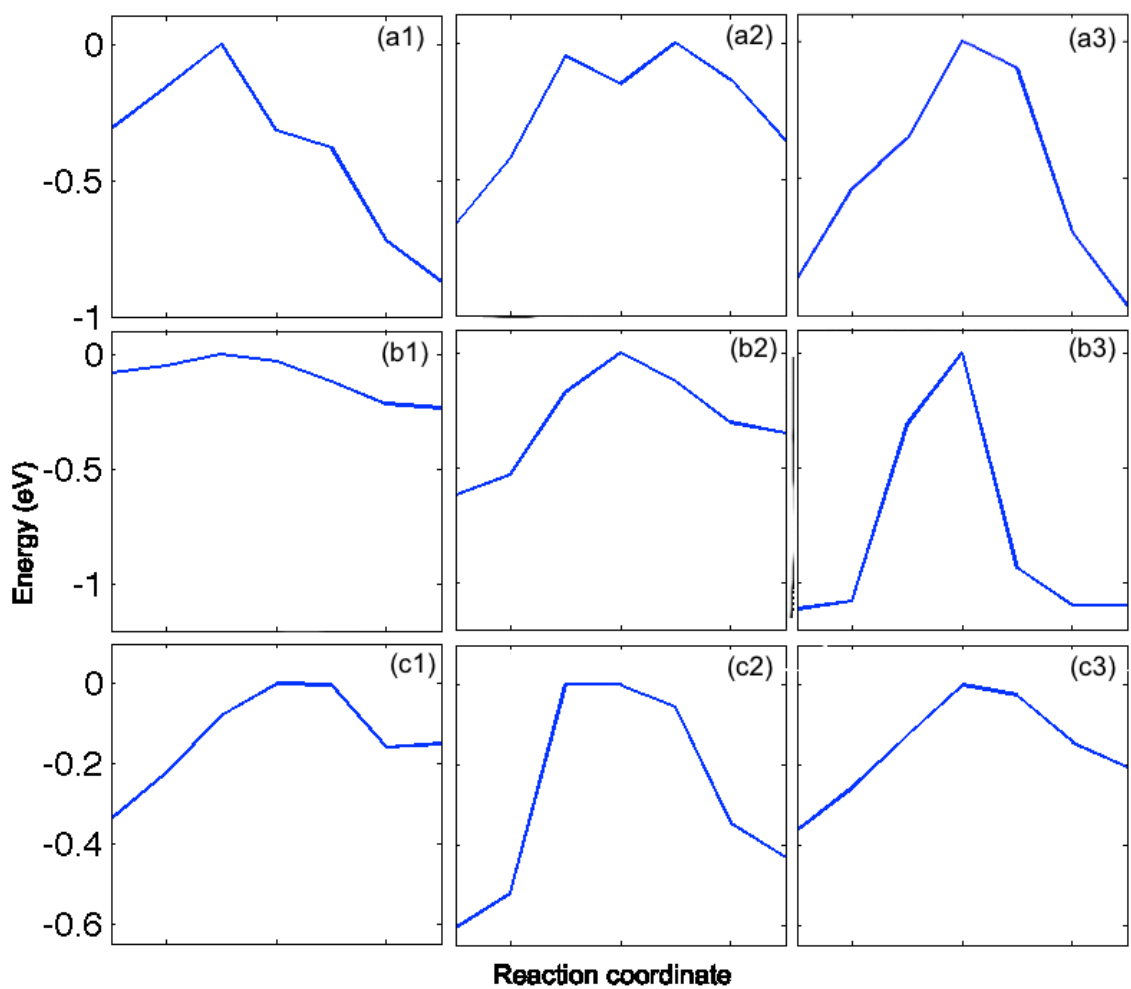

TABLE I

|   | $Li_{1.06}V_3O_8$ (Ref. 62) | $Li_{1.29}V_3O_8$ (Ref. 62) | $L_{1.1}V_3O_8$ (Ref. 22) | $Li_{1.2}V_3O_8$ (Ref. 21) | LDA | LDA+U=2 eV | GGA | GGA+U=3eV |
|---|---|---|---|---|---|---|---|---|
| $a$ | 6.646 | 6.679 | 6.64 | 6.596 | 6.495 | 6.478 | 6.913 | 6.889 |
| $b$ | 3.5928 | 3.607 | 3.59 | 3.559 | 3.527 | 3.554 | 3.583 | 3.625 |
| $c$ | 11.99 | 12.012 | 11.99 | 11.862 | 11.782 | 11.790 | 12.148 | 12.14 |
| $\beta$ | 107.82 | 107.62 | 107.8 | 107.66 | 108.46 | 108.50 | 108.65 | 108.57 |
| $V$ | 272.57 | 275.8 | 271.0 | 265.36 | 256.01 | 257.41 | 285.10 | 287.38 |

TABLE II

|   | $Li_{4.05}V_3O_8$ (Ref. 22) | $Li_4V_3O_8$ (Ref. 21) | LDA | LDA+U=2eV | GGA | GGA+U=3eV |
|---|---|---|---|---|---|---|
| $a$ | 6.03 | 5.955 | 5.838 | 5.894 | 6.0877 | 6.225 |
| $b$ | 3.99 | 3.911 | 3.874 | 3.920 | 3.974 | 4.016 |
| $c$ | 12.2 | 11.915 | 11.622 | 11.809 | 11.983 | 12.329 |
| $\beta$ | 107.5 | 107.03 | 105.41 | 107.22 | 106.39 | 109.74 |
| $V$ | 280.0 | 265.33 | 253.40 | 260.61 | 278.11 | 290.10 |

TABLE III

|  | Clusters | Configurations | M.S.E. | C.V. |
|---|---|---|---|---|
| $\gamma_a$ | 35 | 94 | 0.0454206 | 0.0730165 |
| $\gamma_b$ | 34 | 92 | 0.018001 | 0.0290785 |